\documentclass{article}

\usepackage{arxiv}
\usepackage[english]{babel}
\usepackage{float}
\usepackage[table,xcdraw]{xcolor}
\usepackage{amsmath}
\usepackage{graphicx}
\usepackage{multirow}
\usepackage[colorlinks=true, allcolors=blue]{hyperref}
\usepackage{booktabs}
\begin{document}
\title{Contrastive clustering based on regular equivalence for influential node identification in complex networks}


\author{
 Yanmei Hu\thanks{Corresponding author: Yanmei Hu, E-mail: huyanmei@cdut.edu.cn}\\
  College of Computer and Cyber Security \\
  Chengdu University of Technology \\
  \texttt{huyanmei@cdut.edu.cn} \\
   \And
  Yihang Wu \\
  College of Computer and Cyber Security \\
  Chengdu University of Technology \\
  \texttt{wuyihang@stu.cdut.edu.cn} \\
  \And
  Bing Sun\\
  College of Computer and Cyber Security \\
  Chengdu University of Technology \\
  \texttt{sunbing@stu.cdut.edu.cn}
  \And
  Xue Yue\\
  College of Computer and Cyber Security \\
  Chengdu University of Technology \\
  \texttt{2021050793@stu.cdut.edu.cn}
  \And
  Biao Cai\\
  College of Computer and Cyber Security \\
  Chengdu University of Technology \\
  \texttt{caibiao@cdut.edu.cn} \\
  \And
  Xiangtao Li\\
  School of Artificial Intelligence Jilin University\\
  Jilin University\\
    \texttt{lixt314@jlu.edu.cn} \\
  \And
  Yang Chen\\
  School of Computer Science\\
  Fudan University\\
    \texttt{chenyang@fudan.edu.cn} \\
}

\noindent
This work has been submitted to the IEEE for possible publication. Copyright may be transferred without notice, after which this version may no longer be accessible.
\maketitle

\begin{abstract}
Identifying influential nodes in complex networks is a fundamental task in network analysis with wide-ranging applications across domains. While deep learning has advanced node influence detection, existing supervised approaches remain constrained by their reliance on labeled data, limiting their applicability in real-world scenarios where labels are scarce or unavailable. While contrastive learning demonstrates significant potential for performance enhancement, existing approaches predominantly rely on multiple-embedding generation to construct positive/negative sample pairs. To overcome these limitations, we propose ReCC (\textit{r}egular \textit{e}quivalence-based \textit{c}ontrastive \textit{c}lustering), a novel deep unsupervised framework for influential node identification. We first reformalize influential node identification as a label-free deep clustering problem, then develop a contrastive learning mechanism that leverages regular equivalence-based similarity, which captures structural similarities between nodes beyond local neighborhoods, to generate positive and negative samples. This mechanism is integrated into a graph convolutional network to learn node embeddings that are used to differentiate influential from non-influential nodes. ReCC is pre-trained using network reconstruction loss and fine-tuned with a combined contrastive and clustering loss, with both phases being independent of labeled data. Additionally, ReCC enhances node representations by combining structural metrics with regular equivalence-based similarities. Extensive experiments demonstrate that ReCC outperforms state-of-the-art approaches across several benchmarks.

\end{abstract}

\section{Introduction} \label{introduction}
Complex networks are ubiquitous, with certain nodes playing disproportionately important roles in both structural and functional contexts. These nodes, often referred to as influential nodes, are critical to a wide range of processes such as information diffusion, network robustness, cascading dynamics, and the detection of community structure. Given their importance, the identification of influential nodes has become a central research focus, attracting considerable attention across diverse disciplines. Consequently, numerous methods for identifying such nodes have been proposed.

Traditional methods for identifying influential nodes generally fall into two main categories: metric-based and supervised learning-based methods. Metric-based methods quantify node influence using one or more structural metrics (e.g., degree, betweenness, eigenvector, closeness \cite{hu2022exhaustive}) and greedily select nodes with optimal values as influential nodes; examples include HTS \cite{wan2023novel} and KSGC \cite{yang2021improved}. Supervised learning-based methods frame the identification of influential nodes as a classification or regression problem, leveraging machine learning models, e.g., support vector machine (SVM) and random forest (RF), trained on node features \cite{rezaei2023machine, karoui2022machine}. These features typically derive from structural metrics or, in domain-specific cases, textual data for semantic enrichment \cite{karoui2022machine}. In recent years, deep learning techniques, particularly graph neural networks (GNNs), have been employed to replace traditional machine learning models \cite{zhao2020infgcn, bhattacharya2023detecting}, thereby enhancing the performance of supervised learning-based methods (we refer to these as deep supervised learning methods). Among these, some focus on feature construction: CGCN \cite{zhang2022new} designs features from neighbor-based subnetwork differences, MRCNN \cite{ou2022identification} encodes fixed-size neighborhoods and degrees, and ReGCN \cite{wu2024graph} incorporates regular equivalence similarity. While these methods achieve notable accuracy, their reliance on labeled training data limits applicability in label-scarce scenarios. To address this, we reformalize influential node identification as deep clustering problem, enabling unsupervised resolution.

This reformalized task is a variant of node clustering, where nodes are partitioned into influential and non-influential groups. Given GNN's ability to jointly capture node features and network structure, which makes them ideal for graph-structured data, we adopt a GNN-based clustering paradigm: node features and network structure are fed into a GNN to generate node embeddings, which are then clustered \cite{tsitsulin2023graph}. Recent work integrates contrastive learning to enhance clustering performance by pulling positive sample pairs closer and pushing negative pairs apart in the embedding space. For example, CONVERT \cite{yang2023convert} proposed a reversible perturb-recover network for data augmentation, generating multiple embeddings via a graph convolution network (GCN, a type of GNN) for contrastive learning; Yang et al. \cite{yang2023cluster} and Liu et al. \cite{liu2023hard} employed multi-encoder architectures with cluster-guided strategies to prepare positive/negative pairs; Yang et al. \cite{yang2023dealmvc} encoded cross-view features (global and local) to construct sample pairs. However, these methods rely on generating multiple embeddings through multi-view representations or data augmentation to define positive and negative samples. To overcome this limitation, we propose leveraging regular equivalence-based (RE) similarities to construct sample pairs, eliminating the need for auxiliary embeddings. RE similarity captures structural similarities beyond local neighborhoods, making it particularly suited for assessing influential nodes \cite{wu2024graph}. This advances both unsupervised influential node identification and contrastive learning techniques in network analysis.

Consequently, we propose ReCC (\textit{R}egular \textit{e}quivalence-based \textit{C}ontrastive \textit{C}lustering), a contrastive clustering framework for deep unsupervised influential node identification. ReCC combines GCN with a novel RE similarity-based contrastive learning mechanism (termed ReContrastive) to effectively distinguish influential nodes without labeled data. The framework first employs a GCN to generate node embeddings that incorporate both structural metrics and RE similarity-derived features, capturing both structural properties and global structural equivalence. These embeddings are then optimized through our proposed ReContrastive mechanism, which uniquely generates positive and negative sample pairs directly from RE similarity instead of relying on multiple embeddings like conventional methods. Finally, ReCC is trained through a two-phase process: initial pre-training using network reconstruction loss to preserve topological features, followed by joint fine-tuning with ReConstrastive loss and clustering loss to maximize these separation between influential and non-influential nodes. The main contributions of our work are as follows:

\noindent
\textbf{1) An unsupervised deep clustering model for influential node identification.} In contrast to deep supervised learning methods, we reformalize influential node identification as a deep clustering problem, eliminating label dependence. Further, we propose ReCC, which leverages RE similarity to enable unsupervised learning while maintaining discriminative power.

\noindent
\textbf{2) A contrastive learning mechanism based on RE similarity.} By leveraging RE similarity’s ability to capture similar structural properties beyond neighborhood, we propose ReContrastive, a novel contrastive learning mechanism, to generate positive and negative sample pairs. ReContrastive bypasses the need for multi-embedding generation, simplifying model architecture while outperforming conventional contrastive methods.

\noindent
\textbf{3) Empirical validation on several real datasets.} We conduct comprehensive experiments on several real-world datasets to test ReCC from different aspects. The experimental results show that: (a) ReCC surpasses state-of-the-art clustering methods in accuracy; (b) ReContrastive is pivotal to this performance, demonstrating superior positive/negative sample preparation over traditional contrastive learning; and (c) the integration of structural metrics with RE similarity-derived features enhances ReCC's overall effectiveness. 

\section{Related work}
There are a numerous number of methods for identifying influential nodes. For example, influential spreading models that select nodes with maximizing propagation under specific diffusion dynamics \cite{li2021identification}; influence metric-based methods that apply one or more metrics like degree, betweenness or eigenvector, to evaluate nodes and take nodes with the optimal values as influential nodes \cite{wan2023novel, yang2021improved}; and machine learning-based methods that frame the task as supervised classification or regression problems, and apply machine learning models to solve it \cite{rezaei2023machine, karoui2022machine}. Notably, the machine learning paradigm has evolved from traditional algorithms to contemporary deep learning techniques, particularly deep supervised learning methods. In contrast to these supervised approaches, our ReCC operates as an unsupervised deep clustering framework, fundamentally reformalizing the task as a node clustering problem that groups nodes into influential and non-influential categories. Furthermore, ReCC innovatively incorporates contrastive learning into deep learning-based influential node identification. Accordingly, our related work discussion will focus on two key categories: (1) deep supervised learning methods for influential node identification, and (2) deep node clustering methods, with special attention to those employing contrastive learning techniques.

\textbf{Deep supervised learning methods.} Feature representation through value-encoded vectors constitutes a core component of these methods, with GNNs emerging as the predominant architecture for distinguishing influential nodes. Existing approaches demonstrate diverse feature engineering strategies: Bhattacharya et al. \cite{bhattacharya2023detecting} employed four structural metrics (degree, betweenness, k-shell, and density) as GCN input features. CGCN \cite{zhang2022new} developed a contraction algorithm to compute differences between adjacency matrices of neighbor-derived subnetworks, combining CNN and GNN for processing. MRCNN \cite{ou2022identification} utilized BFS to extract fixed-size neighborhoods, constructing feature matrices from both neighborhood topology and node degrees for CNN classification. InfGCN \cite{zhao2020infgcn} integrated structural metrics (degree, closeness, betweenness, clustering coefficient) with BFS-derived neighbor networks, processed through GCN with Laplacian normalization. RSGNN \cite{fu2023identifying} formulated features using information entropy combined with degree and the average degree of the neighbors, fed into an optimized GCN architecture. RGCN \cite{gao2023key} embedded both local subnetworks and global topology through GCN, and fed the node embeddings to a neural network to score nodes as influential. Kou et al. \cite{kou2023identify} combined graph multi-head attention layers and dense fully connected layers to score nodes by directly taking the adjacency matrix of network as input. 

\textbf{Deep node clustering methods and contrastive learning.} Current methodologies predominantly adhere to a dual-phase framework wherein GNNs first encode both network topology and node features into low-dimensional embeddings, followed by clustering operations performed on the resulting embeddings. Crucially, these phases are typically optimized jointly to enhance clustering performance, as demonstrated in \cite{liu2023dink}. Building upon this foundation, recent advances have increasingly incorporated contrastive learning mechanisms to further improve clustering accuracy. A representative example is CONVERT \cite{yang2023convert}, which introduces a reversible perturb-recover network for data augmentation, generating multiple embeddings through GCN for contrastive learning. Work by Liu et al. \cite{liu2023simple} developed this approach further by employing dual unshared encoders to produce separate embeddings, combining them via k-means clustering while utilizing Gaussian noise perturbation for contrastive similarity measurement. This direction was extended in \cite{liu2023hard} through the introduction of dedicated structure and attribute encoders, where node similarity across different clustering results informed the contrastive learning process, a strategy concurrently explored in \cite{yang2023cluster}.
The field has witnessed significant methodological evolution, particularly through innovations such as NS4GC's \cite{liu2024reliable} learnable similarity matrix for clustering-oriented contrastive learning and MAGI's \cite{liu2024revisiting} augmentation-free reformulation of modularity maximization using random walks. Multi-view approaches have similarly advanced, exemplified by \cite{yang2023dealmvc}'s attention-based fusion of cross-view features and \cite{jin2023deep}'s cross-view completion strategy for handling incomplete data. Parallel developments include subgraph-based embedding techniques \cite{dong2024subgraph}, GAN-augmented multi-view alignment \cite{yang2024trustworthy}, and specialized solutions for attribute-missing graphs \cite{tu2024attribute}. In spite of these advances, three fundamental limitations persist: First, the prevailing dependence on multi-embedding generation for contrastive sample preparation; second, the spatial constraints imposed by neighborhood-based methods like MAGI's random walk approach, which seems suboptimal for capturing relationships between distant influential nodes; and third, the absence of mechanisms explicitly leveraging global structural equivalence. Our ReCC framework addresses these limitations through its innovative integration of RE similarity, eliminating the need for multiple embeddings while enabling global structural alignment.

\section{Methodology} \label{method}
\subsection{Preliminaries} \label{Pre}
\textbf{Problem statement}.\label{subsec:ps} This study aims to identify influential nodes in a network $G=(V,E)$ solely based on its network structure using a deep unsupervised learning approach. Here, $V$ denotes the set of nodes and $E$ the set of edges connecting them. Formally, this unsupervised task is defined as the following deep clustering learning problem:
\begin{equation}
    \{C_1, C_2\}=f(G|\boldsymbol{\Theta})
    \label{eq:ps}
\end{equation}
where $C_1$ and $C_2$ represent the sets of influential and non-influential nodes, respectively, and $f$ is the deep clustering model with learnable parameters $\boldsymbol{\Theta}$. To address this problem, we introduce ReCC, a framework that significantly deviates from conventional influential node identification methods.  

\textbf{Regular equivalence-based similarity}.\label{subsec:re} RE similarity is crucial for ReCC, so we briefly describe it before introducing ReCC. For a pair of nodes $v_i$ and $v_j$ in $G$, their RE similarity is defined as follows:
\begin{equation}
    \boldsymbol{S}^{re}(i, j) = \alpha \sum_{k,l}{\boldsymbol{A}_{i,k}\boldsymbol{A}_{j,l}S^{re}(k,l)}
    \label{eq:sre1}
\end{equation}
where $\boldsymbol{A}$ denotes the adjacency matrix of $G$, and $\alpha$ is a parameter within the range $(0, 1/\lambda_{max})$ ($\lambda$ is the largest eigenvalue of $\boldsymbol{A}$). The calculation of $\boldsymbol{S}^{re}(i,j)$ is self-referential because it requires $\boldsymbol{S}^{re}(k,l)$ for neighbors $v_k$ and $v_l$ of $v_i$ and $v_j$, respectively. To address this, Eq. (\ref{eq:sre1}) is relaxed to:
\begin{equation}
    \boldsymbol{S}^{re}(i,j) = \alpha \sum_{k}{\boldsymbol{A_{i,k}}\boldsymbol{S}^{re}(k,j)}
    \label{eq:sre2}
\end{equation}
This implies that $v_j$ is regular equivalence similar to $v_i$ if $v_j$ is regular equivalence similar to the neighbors of $v_i$. By recursively calculating Eq. (\ref{eq:sre2}) starting from $\boldsymbol{S}^{re}(i,i) = 1$ ($i=1,2,...,|V|$), the RE similarities between nodes can be obtained. Additionally, Eq. (\ref{eq:sre2}) can be transformed into matrix formulation, allowing RE similarity to be calculated using matrix operations. More details are referred to \cite{wu2024graph}.

From Eq. (\ref{eq:sre2}), it is evident that two nodes are regular equivalence similar if they have many similar neighbors, which differs from the commonly used Jaccard and Cosine similarities that require many common neighbors. RE similarity captures the structural equivalence of nodes without common neighbors, making it suitable for measuring the similarity between influential nodes that may not be directly connected or have no common neighbors. Motivated by this, ReCC integrates RE similarity-derived features into its feature construction process and develops a RE similarity-based contrastive learning mechanism.

\subsection{Framework} \label{subsec:framework}
\begin{figure*}
    \centering
    \includegraphics[width=\textwidth]{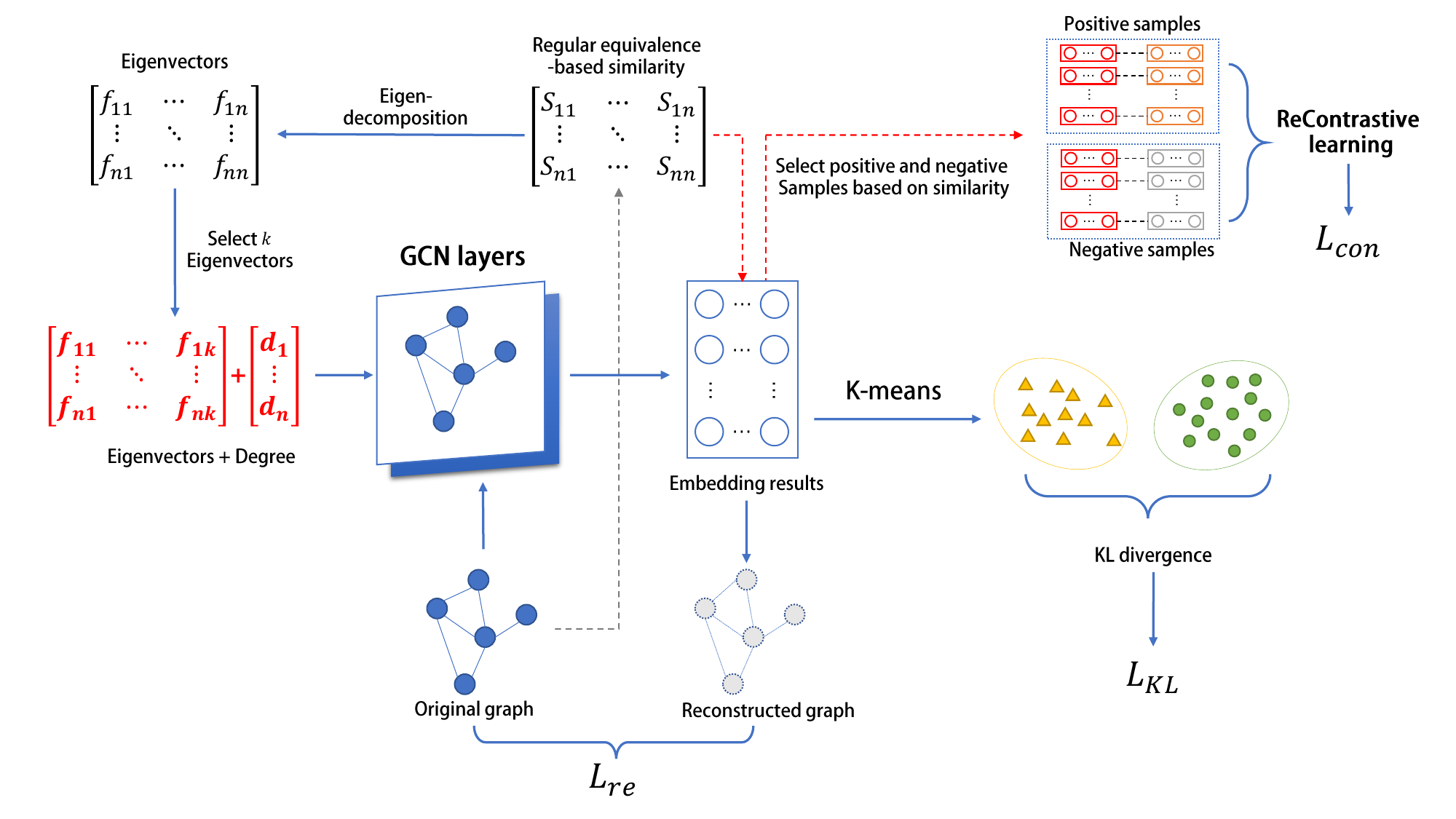}
    \caption{The architecture of ReCC.}
    \label{fig:frame}
\end{figure*}

Fig. \ref{fig:frame} illustrates the overall architecture of ReCC. The framework operates through four key components: 1) feature construction combining RE similarity-derived features and structural metrics, particularly node degree (Section \ref{subsec:feature-construction}), 2) node embedding generation using both the constructed features and network adjacency matrix (Section \ref{subsec:gcn-embedding}), 3) clustering to separate influential from non-influential nodes, and 4) the proposed ReContrastive that enhances embeddings through RE similarity-based contrastive learning (Section \ref{subsubsec:ReContranstive}). The complete model is optimized in a self-supervised manner with reconstruction loss, ReContrastive loss, and clustering loss.

\subsection{Feature construction} \label{subsec:feature-construction}
Structural metrics, which capture structural position from either global (e.g., Betweenness and PageRank \cite{hu2022exhaustive}) or local (e.g., degree, egonet density, and local clustering coefficient \cite{hu2022exhaustive}) perspectives, can serve as node features (note that these mentioned structural metrics are also called centrality metrics in the literature). Further, as discussed in Section \ref{subsec:re}, RE-based similarity effectively identifies structural similarities between influential nodes, even when they are not directly connected or lack common neighbors. To enhance node representation, we combine structural metrics with RE-based similarity. Specifically, correlation analysis of structural metrics reveals significant redundancy despite differing definitions (see Section \ref{sec:appendix}). Thus, we focus on representative metrics, and further select \textit{degree} due to its demonstrated efficiency and effectiveness (see Section \ref{subsubsec:ablation-features}). Regarding RE-based similarity, we derive features from pairwise node similarities. Since a single similarity value only reflects the structural resemblance between two nodes and is insufficient as a standalone feature, we instead consider the similarity profile of each node with respect to all others. If two nodes exhibit similar similarity distributions, they likely occupy analogous structural roles. However, using all pairwise similarities leads to high-dimensional and noisy features. To address this, we first calculate the RE-based similarity matrix $\boldsymbol{S}^{re}$, then extract its most informative eigenvectors, denoted as $\boldsymbol{ReEig}$, as low-dimensional node features. The final feature matrix is constructed as follows:
\begin{equation}
        \boldsymbol{X} = [\boldsymbol{ReEig},\boldsymbol{d}],
\end{equation}
where $\boldsymbol{d}$ is a column vector whose elements represent node degrees. To identify the most informative eigenvectors, we adopt the method proposed in \cite{wu2024graph}: we rank the eigenvalues of $\boldsymbol{S}_{re}$ in descending order and analyze the resulting eigenvalue curve. The eigenvectors corresponding to eigenvalues before the first significant drop form $\boldsymbol{ReEig}$.

\subsection{Graph convolution network embedding} \label{subsec:gcn-embedding}
GCN has emerged as a powerful and efficient deep learning approach for graph-structured data, particularly due to their ability to effectively encode both node features and network structure into meaningful latent representations. These representations prove crucial for various downstream tasks. Based on these strengths, we employ a GCN in ReCC to generate high-quality node embeddings.

The GCN architecture processes the input graph $G$ and feature matrix $\boldsymbol{X}$, producing node embeddings through multiple GCN layers. Each layer aggregates information from neighboring nodes to refine the representations. Formally, the hidden representation of node $v_i$ at layer $l$ is calculated as:   
\begin{equation}
\begin{aligned}
\boldsymbol{H}^{l}_i  = \Phi(\sum_{v_j \in N(v_i) \cap \{v_i\}} \frac {1} {\sqrt{\boldsymbol{d}(i)} \cdot \sqrt{\boldsymbol{d}(j)}} \\ \cdot
(\boldsymbol{\Theta_{c}}^l)^T \cdot \boldsymbol{H}^{l-1}_{j}),\label{eq:GCN_agg}
\end{aligned}
\end{equation}
where $N(v_i)$ denotes the set of neighboring nodes for $v_i$, $\boldsymbol{\Theta}_{c}^l$ represents the trainable weight matrix at layer $l$, $\Phi$ is the Exponential Linear Unit (ELU) activation function, and  $\boldsymbol{H}^0_i = \boldsymbol{X}_i$ serves as the initial node representation.

\subsection{Self-optimizing graph embedded clustering with ReContrastive}\label{subsec:soc} 
Following the GCN processing, ReCC performs node embedding clustering to differentiate between influential and non-influential nodes. Obtaining optimal node embeddings for this clustering task is therefore crucial to the model's performance. Our training approach consists of two phases: 1) an initial pre-training phase focused on network structure reconstruction, followed by 2) a fine-tuning phase that incorporates both RE contrastive learning and minimization of the Kullback–Leibler (KL) divergence between predicted and target cluster distributions.

\subsubsection{Pre-training with network reconstruction}\label{subsubsec:pretraining}
The pre-training phase aims to learn task-agnostic node embeddings that effectively encode both node features and network structure information. In this phase, we optimize the model by minimizing the network reconstruction loss:
\begin{equation}
    L_{re} = ||\mathbf{A} - \hat{\mathbf{A}} ||_{2} \label{eq:loss_re}
\end{equation}
where $\hat{\mathbf{A}} =  \sigma(\mathbf{Z}^T \mathbf{Z})$ represents the reconstructed adjacency matrix through a sigmoid activation function $\sigma(\cdot)$, and $\boldsymbol{Z}$ denotes the learned node embedding matrix.

\subsubsection{Regular equivalence-based contrastive learning}\label{subsubsec:ReContranstive}
The effectiveness of contrastive learning depends fundamentally on the principled preparation of positive and negative samples. Conventional approaches typically generate multiple embeddings through input augmentation (e.g., noise injection) or exploit inherently available multi-view data representations. In these approaches, two principal strategies are employed for sample preparation: 1) different embeddings derived from the same instance constitute positive pairs, with all other combinations treated as negative pairs; or 2) instances identified as similar through their multiple embeddings form positive pairs, while dissimilar instances are designated as negative pairs. In contrast, we propose ReContrastive, an innovative paradigm that determines positive and negative samples through RE similarity. Specifically, for each node $v_i$, we define:
\begin{itemize}
    \item Positive samples ($pos_i$): the $k_p$ nodes exhibiting highest RE similarity to $v_i$.
    \item Negative samples ($neg_{i}$): the $ k_n$ nodes exhibiting lowest RE similarity to $v_i$.
\end{itemize} 
To determine $v_i$'s positive and negative samples, we sort all other nodes by their RE similarities (from the $i$-th row of $\boldsymbol{S^{re}}$), then select top $k_p$ nodes as $pos_i$ and bottom $k_n$ nodes as $neg_i$. With the sample sets determined, the contrastive loss is formulated as follows:
\begin{equation}
    L_{con} = \frac{1}{N} \sum_{i=1}^{N} {l_i},
    \label{eq:con_loss}
\end{equation}
\begin{equation}
    l_i = -\log \frac{\sum_{j \in \text{pos}_{i}} e^{\cos(\mathbf{Z}_i,\mathbf{Z}_j)}}{\sum_{k \in \text{neg}_i} e^{\cos(\mathbf{Z}_i,\mathbf{Z}_k)}}
    \label{eq:sig_con_loss}
\end{equation}
where $cos(.)$ denotes Cosine similarity. Notably, the positive ($pos_i$) and negative ($neg_i$) sample sets for each node require only a single determination following computation of the RE similarity matrix $\boldsymbol{S}_{re}$. We intentionally set small values for $k_p$ and $k_n$, selecting only the most similar nodes as reliable positive samples and the most dissimilar nodes as confident negative samples. This design yields three key advantages: 1) computational efficiency during learning iterations, as Eq. (\ref{eq:sig_con_loss}) involves only limited terms in both numerator and denominator; 2) elimination of multi-embedding generation requirements; and 3) guaranteed sample quality through extreme similarity selection.

\subsubsection{KL-divergence}\label{subsubsec:kl}
To further optimize the clustering quality, we incorporate a KL-divergence objective between predicted and target cluster distributions during model fine-tuning. Formally, given the node embeddings $\boldsymbol{Z}$, we compute the predicted cluster distribution as follows: 
\begin{equation}
    q_{ij} = \frac{(1 + ||\mathbf{Z}_i - \mathbf{c}_j||^2 / \tau)^{-\frac{\tau + 1}{2}}}{\sum_{j'}(1 + ||\mathbf{Z}_i - \mathbf{c}_{j'}||^2 / \tau)^{-\frac{\tau + 1}{2}}},
\end{equation}
where $q_{ij}$ represents the soft assignment probability of node $v_i$ to cluster $j$, $\boldsymbol{c}_j$ is the centroid embedding of cluster $j$ (obtained by averaging all node embeddings assigned to the cluster), and $\tau$ serves as temperature parameter controlling distribution sharpness (default to 1). Using these soft assignments, we construct the target distribution $p_{ij}$ as:
\begin{equation}
    p_{ij} = \frac{q_{ij}^2 / f_j}{\sum_{j'} q_{ij'}^2 / f_{j'}},
\end{equation}
where $f_j = \sum_i q_{ij}$ denotes the soft cluster frequency for cluster $j$, obtained by aggregating assignment probabilities across all nodes. The resulting KL-divergence loss between the target distribution $p_{ij}$ and predicted distribution $q_{ij}$ is then given by:
\begin{equation}
    L_{KL} = \sum_{i}\sum_{j}p_{ij}\log \frac{p_{ij}}{q_{ij}}.
    \label{KL-eq}
\end{equation}
Through the minimization of this KL-divergence objective, the model learns to align the predicted cluster assignments $q_{ij}$ with the target distribution $p_{ij}$. This optimization process effectively drives the node embeddings to organize into clusters that increasingly resemble the target grouping structure.

\subsubsection{Fine-tuning: joint embedding and clustering optimization}\label{subsubsec:fine-tuning}
Following the pre-training phase, ReCC undergoes fine-tuning through a combined optimization objective that integrates both the ReContrastive loss and KL-divergence loss. The fine-tuning objective is formulated as:
\begin{equation}
    L =  L_{con} +  L_{KL}
\label{eq:total_loss}
\end{equation}
During fine-tuning, we minimize the total loss (Eq. (\ref{eq:total_loss})) through backpropagation with the Adam optimizer. This iterative gradient-based optimization simultaneously refines node embeddings to better capture structural properties and improves cluster quality. The gradient computation for the ReContrastive loss ($L_{con}$) component with respect to node embedding $\boldsymbol{Z}_i$ is:
\begin{equation}
    \frac{\partial L_{con}}{\partial \mathbf{Z}_i} = \frac{1}{N} \sum_{i=1}^{N} \frac{\partial l_i}{\partial \mathbf{Z}_i},
\end{equation}
\begin{align}
\frac{\partial l_i}{\partial \mathbf{Z}_i}
&=
-\sum_{j \in \text{pos}_i}
\frac{e^{\cos(\mathbf{Z}_i, \mathbf{Z}_j)}}
     {\sum_{j\in \text{pos}_{i}} e^{\cos(\mathbf{Z}_i,\mathbf{Z}_{j})}}
\frac{\mathbf{Z}_j}{\|\mathbf{Z}_i\|\|\mathbf{Z}_j\|} \notag\\[2pt]
&\quad
+\sum_{j \in \text{pos}_i}
\frac{e^{\cos(\mathbf{Z}_i, \mathbf{Z}_j)}}
     {\sum_{j \in \text{pos}_{i}} e^{\cos(\mathbf{Z}_i,\mathbf{Z}_{j})}}
\cos(\mathbf{Z}_i, \mathbf{Z}_j)\,
\frac{\mathbf{Z}_i}{\|\mathbf{Z}_i\|^2} \notag\\[2pt]
&\quad
+\sum_{k \in \text{neg}_i}
\frac{e^{\cos(\mathbf{Z}_i, \mathbf{Z}_k)}}
     {\sum_{k \in \text{neg}_{i}} e^{\cos(\mathbf{Z}_i,\mathbf{Z}_{k})}}
\frac{\mathbf{Z}_k}{\|\mathbf{Z}_i\|\|\mathbf{Z}_k\|} \notag\\[2pt]
&\quad
-\sum_{k \in \text{neg}_i}
\frac{e^{\cos(\mathbf{Z}_i, \mathbf{Z}_k)}}
     {\sum_{k\in \text{neg}_{i}} e^{\cos(\mathbf{Z}_i,\mathbf{Z}_{k})}}
\cos(\mathbf{Z}_i, \mathbf{Z}_k)\,
\frac{\mathbf{Z}_i}{\|\mathbf{Z}_i\|^2}
\end{align}
Meanwhile, the gradient of the KL-divergence loss with respect to node embedding $\boldsymbol{Z}_i$ is computed as:
\begin{align}
\frac{\partial L_{KL}}{\partial \mathbf{Z}_i}
&= \frac{\alpha + 1}{\alpha} \sum_{j}
   \left( 1 + \frac{\|\mathbf{Z}_i - \mathbf{c}_j\|^2}{\alpha} \right)^{-1}
   \times \notag \\
&\quad (p_{ij} - q_{ij})(\mathbf{Z}_i - \mathbf{c}_j).
\end{align}
The complete gradient with respect to node embedding $\mathbf{Z}_i$ combines both loss components, i.e.,  $\frac{\partial L_{total}}{\partial \mathbf{Z}_i} = \frac{\partial L_{con}}{\partial \mathbf{Z}_i} + \frac{\partial L_{KL}}{\partial \mathbf{Z}_i}$. This aggregated gradient is then backpropagated through the GCN architecture. Using standard backpropagation, we compute the gradient with respect to the model parameters ($\boldsymbol{\Theta}_c$) at iteration $t$: $\boldsymbol{g_t} = \frac{\partial L}{\partial \boldsymbol{\Theta}_c}$. The first-order momentum of this gradient, $\boldsymbol{m}_t$, and the second-order momentum of the square of this gradient, $\boldsymbol{v}_t$, are computed as:   
\begin{equation}
    \boldsymbol{m}_t = \beta_1 \boldsymbol{m}_{t-1}+(1-\beta_1)\boldsymbol{g}_t,
\end{equation}
\begin{equation}
    \boldsymbol{v}_t = \beta_2 \boldsymbol{v}_{t-1} +(1-\beta_2)\boldsymbol{g}_{t}^{2},
\end{equation}
where $\beta_1$ and $\beta_2$ are exponential decay rates for the momentum estimates. Furthermore, to avoid the influence of bias, we compute: $\hat{\boldsymbol{m}_t}=\frac{\boldsymbol{m}_t}{1-\beta^{t}_1}$ and $ \hat{\boldsymbol{v}_t}=\frac{\boldsymbol{v}_t}{1-\beta^{t}_2}$. Finally, $\boldsymbol{\Theta}_{c,t}$ is updated as follows:
\begin{equation}
    \boldsymbol{\Theta}_{c,t} = \boldsymbol{\Theta}_{c,t-1} - l_r \frac{\hat{\boldsymbol{m}_t}}{\sqrt{\hat{\boldsymbol{v}_t}}+\epsilon},
\end{equation}
where $l_r$ is the learning rate, and $\epsilon$ is a numerical stability constant that prevents division by zero in the parameter update.

\section{Experiment}

\subsection{Experimental setting} \label{settings}
We evaluate ReCC against 7 baseline methods on 8 public datasets: ca-HepTh \cite{leskovec2007graph}, Facebook \cite{mcauley2012social}, Crime, Erd\H{o}s, Euroroad, Hamsterster, Sister cities, and Wikipedia \cite{kunegis2013konect}. The baselines include traditional methods (NMF+KM and LCCF \cite{cai2010locally}), deep learning methods (DEC \cite{xie2016unsupervised} and DCN \cite{yang2017towards}), and contrastive clustering methods (CONVERT \cite{yang2023convert}, NS4GC \cite{liu2024reliable}, and MAGI \cite{liu2024revisiting}). NMF+KM is a clustering method based on Non-negative Matrix Factorization (NMF) and k-means. LCCF is based on NMF with a graph Laplacian regularization and simultaneously optimizes feature matrix, latent semantic factor matrix and membership matrix to obtain embeddings, based on which k-means is used for clustering. DEC is a method that uses a stack autoencoder (SAE) for embedding and conduct clustering based on embeddings. DCN follows a similar architecture to DEC, but employs a least-squares loss instead. CONVERT generates an additional embedding via data augmentation, treating the same instance across different embeddings as positive and all others as negative. NS4GC creates two augmented embeddings but uses their pairwise similarity to define positive and negative samples. In contrast, MAGI relies on random walks to construct its sample pairs. 

All baselines are configured as their original papers. ReCC employs a 3-layer GCN (128 units per layer), pre-trained for 100 epoches and fine-tuned for another 100 epoches (with dropout). $k_p$ and $k_n$ are set to 2 and 1, respectively. Similar to conventional methods, k-means is employed to perform clustering on node embeddings in ReCC. All experiments are conducted in PyTorch on a PC with Intel(R) Core(TM) i9-10850K CPU, Nvidia GTX 1650 GPU, and 32GB RAM. The implementation of ReCC can be accessed from github \footnote{https://github.com/Huyanmei123/ReCC}.

\begin{table*}[h]
\caption{Identification performance of different methods.}
\centering
\resizebox{\textwidth}{!}{%
\begin{tabular}{cccccccccc}
\hline
Methods                                                                         & Metrics & ca-HepTh        & Facebook        & Crime           & Erd\H{o}s       & Euroroad        & Hamsterster     & Sister cities   & Wikipedia       \\ \hline
\multirow{3}{*}{NF+KM}& ACC     & 0.1449          & 0.1700          & 0.1306          & 0.1365          & 0.1325          & 0.1543          & 0.1333          & 0.1910          \\
                                                                                & NMI     & 0.0381          & 0.1023          & 0.0158          & 0.0192          & 0.0173          & 0.0582          & 0.0000          & 0.1550          \\
                                                                                & ARI     & 0.0345          & 0.1319          & 0.0238          & 0.0147          & 0.0257          & 0.0694          & 0.0000          & 0.1694          \\ \hline
\multirow{3}{*}{LCCF}                                                           & ACC     & 0.0937          & 0.0815          & 0.0711          & 0.0768          & 0.0490          & 0.0876          & 0.1333          & 0.0781          \\
                                                                                & NMI     & 0.0233          & 0.0300          & 0.0348          & 0.0303          & 0.0385          & 0.0303          & 0.0000          & 0.0285          \\
                                                                                & ARI     & -0.0056         & 0.0031          & -0.0071         & 0.0083          & -0.0014         & -0.0024         & 0.0000          & -0.0036         \\ \hline
\multirow{3}{*}{DEC}                                                            & ACC     & 0.1333          & 0.1331          & 0.1322          & 0.1332          & 0.1325          & 0.1333          & 0.1415          & 0.1331          \\
                                                                                & NMI     & 0.0000          & 0.0000          & 0.0000          & 0.0000          & 0.0000          & 0.0000          & 0.0394          & 0.0000          \\
                                                                                & ARI     & 0.0000          & 0.0000          & 0.0000          & 0.0000          & 0.0000          & 0.0000          & 0.0390          & 0.0000          \\ \hline
\multirow{3}{*}{DCN}                                                            & ACC     & 0.1333          & 0.1331          & 0.1322          & 0.1332          & 0.1325          & 0.1333          & 0.0870          & 0.1331          \\
                                                                                & NMI     & 0.0000          & 0.0000          & 0.0000          & 0.0000          & 0.0000          & 0.0000          & 0.0230          & 0.0000          \\
                                                                                & ARI     & 0.0000          & 0.0000          & 0.0000          & 0.0000          & 0.0000          & 0.0000          & -0.0082         & 0.0000          \\ \hline
\multirow{3}{*}{CONVERT}                                                        & ACC     & 0.6933          & 0.9424          & 0.6950          & 0.7305          & 0.7007          & 0.9543          & 0.7950          & 0.9604          \\
                                                                                & NMI     & 0.0582          & 0.6688          & 0.0630          & 0.1083          & 0.0665          & 0.7214          & 0.3919          & 0.7872          \\
                                                                                & ARI     & 0.0980          & 0.7792          & 0.0944          & 0.1624          & 0.0912          & 0.8233          & 0.3492          & 0.8467          \\ \hline
\multirow{3}{*}{MAGI} & ACC & 0.6443 & 0.6606 & 0.5289 & 0.6178 & 0.6490 & 0.5843 & 0.7636 & 0.6655 \\
                      & NMI & 0.2535 & 0.2683 & 0.1593 & 0.2305 & 0.0125 & 0.0765 & 0.3786 & 0.2729 \\
                      & ARI & 0.0679 & 0.0899 & -0.0431 & 0.0349 & 0.0422 & -0.0544 & 0.2759 & 0.0969 \\ \hline
\multirow{3}{*}{NS4GC} & ACC & 0.8812 & 0.8136 & 0.7355 & 0.7587 & 0.7301 & 0.7378 & 0.7823 & 0.9245 \\
                       & NMI & 0.4676 & 0.4346 & 0.3456 & 0.1513 & 0.1915 & 0.3052 & 0.4029 & 0.6718 \\
                       & ARI & 0.5786 & 0.3925 & 0.2128 & 0.2509 & 0.2071 & 0.2215 & 0.3176 & 0.7192 \\ \hline
\multirow{3}{*}{ReCC}                                                           & ACC     & \textbf{0.9704}& \textbf{0.9401}& \textbf{0.9488} & \textbf{0.8847} & \textbf{0.7828} & \textbf{0.9616}& \textbf{0.9868} & \textbf{0.9694} \\
                                                                         & NMI     & \textbf{0.8267}& \textbf{0.7181}& \textbf{0.7333} & \textbf{0.5135} & \textbf{0.2774} & \textbf{0.7921}& \textbf{0.9058} & \textbf{0.8229} \\
                                                                                & ARI     & \textbf{0.8840}& \textbf{0.7735}& \textbf{0.8032} & \textbf{0.5809} & \textbf{0.3156} & \textbf{0.8509}& \textbf{0.9474} & \textbf{0.8803} \\ \hline

\end{tabular}%
}

\label{tab:compared-results}
\end{table*}

\subsection{Comparison with other methods}\label{subsec:exp-result}
\subsubsection{Performance comparison}
We assess model performance using three metrics: accuracy (ACC), normalized mutual information (NMI), and adjusted Rand index (ARI). As shown in Table \ref{tab:compared-results}, ReCC demonstrates strong performance across all evaluation metrics, achieving: ACC value exceeding 0.9, NMI exceeding 0.7, and ARI exceeding 0.75 on 6 out of 8 datasets. Moreover, ReCC outperforms all baselines across all datasets. The traditional methods (NMF+KM and LCCF) perform poorly, failing to meaningfully identify influential nodes. The deep learning methods (DEC and DCN) are entirely ineffective, with near-zero NMI and ARI scores. While contrastive clustering methods (CONVERT, MAGI, and NS4GC) show significant improvement over traditional and deep learning methods, they still fall short of ReCC's performance.

These results demonstrate that ReCC achieves state-of-the-art performance in unsupervised influential node identification, surpassing existing clustering methods - including the most recent contrastive clustering approaches (CONVERT, MAGI, and NS4GC) - across multiple evaluation metrics.

\subsubsection{Efficiency comparison of different contrastive learning mechanisms}
We also compare the running time required for contrastive learning loss computation across different methods. As shown in Table \ref{tab:contrastive-loss-time}, which reports the average running time per epoch for each method (methods without contrastive learning are excluded), ReCC consistently requires the least amount of time for contrastive loss calculation across all datasets, followed by MAGI and NS4GC, while CONVERT consumes the most time.

CONVERT adopts the most conventional contrastive learning approach: it generates multiple embeddings (typically two) and treats the same node across different embedding views as a positive pair, while considering all other nodes as negative samples (details regarding the formulation of contrastive losses in the compared methods can be found in their respective original papers). This leads to a computationally expensive contrastive loss. NS4GC also generates two embeddings, but defines its contrastive loss based on the similarity matrix between them. Specifically, the loss consists of the diagonal entries of the similarity matrix, the sum of similarities over connected nodes, and an L1-norm sparsity penalty applied to similarities of unconnected nodes. Although NS4GC also implicitly incorporates all other nodes in the contrastive loss for each node, it is more efficient than CONVERT, which probably is due to the absence of logarithmic operation. In contrast to these methods, both MAGI and ReCC eliminate the need for generating multiple embeddings. MAGI uses a modularity matrix approximated via random walks to identify positive and negative pairs: nodes within the same approximated sub-community form positive pairs, while those from different sub-communities form negative pairs. Because the sampling is constrained within local random-walk neighborhoods, the set of positive and negative samples is significantly smaller compared to conventional contrastive learning approaches. Nevertheless, it remains larger than that in ReCC. In ReCC, positive and negative samples for each node are determined based on RE similarity: the most RE-similar nodes form positive pairs, while the most RE-dissimilar nodes form negative pairs (see Eq. (\ref{eq:con_loss})). This results in very few samples per node (specifically, 2 positive and 1 negative sample), making contrastive loss computation highly efficient compared to existing graph contrastive learning methods.


\begin{table*}[h]
\centering
\caption{The average time consumption of contrastive learning loss per epoch in different methods (seconds)}
\label{tab:contrastive-loss-time}
\resizebox{\linewidth}{!}{
\begin{tabular}{ccccccccc}
\hline
Methods & ca-HepTh & Facebook & Crime & Erd\H{o}s&
  Euroroad &
  Hamsterster &
  Sister cities &
  Wikipedia \\
\hline
CONVERT & 0.6022 & 0.1703 & 0.0152 & 0.4367 & 0.0231 & 0.0475 & 1.2601 & 0.1444 \\
NS4GC &  0.0305 & 0.0215 & 0.0019 & 0.0200 & 0.0020 & 0.0045 & 0.0491 & 0.0186 \\
MAGI &  0.0067 & 0.0030 & 0.0016 & 0.0054 & 0.0016 & 0.0014 & 0.0095 & 0.0025 \\
ReCC & \textbf{0.0010} & \textbf{0.0009} & \textbf{0.0011} & \textbf{0.0009} & \textbf{0.0010} & \textbf{0.0009} & \textbf{0.0012} &\textbf{ 0.0010} \\
\hline
\end{tabular}
}
\end{table*}

\subsection{Ablation study}
\subsubsection{Testing of losses} \label{subsubsec:ablation-loss}
To evaluate the contributions of the ReConstrastive loss and KL-divergence loss in ReCC, we compare two ablated variants, which are ReCC(ReContrastive) that excludes KL-divergence loss and  ReCC(KL) that excludes ReConstrastive loss. As shown in Table \ref{tab:test_loss}, while the performance differences between ReCC and its variants appear marginal, both variants exhibit consistent declines: ReCC(ReContrastive) underperforms the full model on 5/8 datasets, and ReCC(KL) underperforms on 7/8 datasets. This demonstrates that the combined use of both losses is optimal; removing either component leads to measurable performance degradation.

\begin{table*}[h]
\caption{Testing of different losses.}
\centering
\resizebox{\textwidth}{!}{%
\begin{tabular}{cccccccccc}
\hline
Models                                                                          & Metrics & ca-HepTh        & Facebook        & Crime           & Erd\H{o}s       & Euroroad        & Hamsterster     & Sister cities   & Wikipedia       \\ \hline
\multirow{3}{*}{ReCC}                                                                            & ACC     & 0.9704          & 0.9401          & \textbf{0.9488} & \textbf{0.8847} & \textbf{0.7828} & 0.9616          & \textbf{0.9868} & \textbf{0.9694} \\
                                                                                & NMI     & 0.8267          & 0.7181          & \textbf{0.7333} & \textbf{0.5135} & \textbf{0.2774} & 0.7921          & \textbf{0.9058} & \textbf{0.8229} \\
                                                                                & ARI     & 0.8840          & 0.7735          & \textbf{0.8032} & \textbf{0.5809} & \textbf{0.3156} & 0.8509          & \textbf{0.9474} & \textbf{0.8803} \\ \hline
\multirow{3}{*}{ReCC (ReContrastive)}& ACC     & \textbf{0.9718} & \textbf{0.9434} & 0.9421          & 0.8610          & 0.7814          & \textbf{0.9625} & 0.9823          & 0.9604          \\
                                                                                & NMI     & \textbf{0.8286} & \textbf{0.7281} & 0.7249          & 0.4427          & 0.2757          & \textbf{0.7951} & 0.8735          & 0.7872          \\
                                                                                & ARI     & \textbf{0.8892} & \textbf{0.7851} & 0.7794          & 0.5019          & 0.3126          & \textbf{0.8543} & 0.9298          & 0.8467          \\ \hline
\multirow{3}{*}{ReCC(KL)}                                                       & ACC     & 0.9691          & 0.9401          & 0.9442          & 0.8784          & 0.7814          & 0.9603          & \textbf{0.9868} & 0.9604          \\
                                                                                & NMI     & 0.8215          & 0.7176          & 0.7230          & 0.4920          & 0.2757          & 0.7874          & \textbf{0.9058} & 0.7872          \\
                                                                                & ARI     & 0.8793          & 0.7733          & 0.7868          & 0.5579          & 0.3126          & 0.8461          & \textbf{0.9474} & 0.8467          \\ \hline
\end{tabular}%
}

\label{tab:test_loss}
\end{table*}

\subsubsection{Testing of different features} \label{subsubsec:ablation-features}
As described in Section \ref{subsec:feature-construction}, we initially selected only degree among numerous structural metrics and combined it with RE similarity-derived features for feature construction. Here we further evaluate alternative combinations of structural metrics.

Correlation analysis revealed high redundancy among structural metrics. We therefore select representative metrics: 1) Eigenvector and Pagerank as representatives of global structural metrics, and 2) degree, extended degree, density of egonet, conductance of egonet, local clustering coefficient, and core dominance pearson as representatives of local structural metrics (see Section \ref{sec:appendix} for details). We then test seven feature combinations: RE similarity + degree ("ReEig+Degree"), degree alone ("Degree"), RE similarity + all representative metrics ("ReEig+All"), all representative metrics alone ("All"), RE similarity + local representative metrics ("ReEig+Local"), local representative metrics alone ("Local"), and RE similarity alone ("ReEig"). Table \ref{tab:test_fea} shows the results of different feature combinations. "ReEig+Degree" achieves the best performance, while all other feature combinations exhibit declines: "Degree" declines on 6/8 datasets, "ReEig+ALL" on 5/8, "All" on 6/8, "ReEig+Local" on 2/8, "Local" on 3/8, and "ReEig" on 8/8. Omitting RE similarity-derived features consistently reduces performance: "ALL" declines on 4/8 datasets, "Local" on 3/8, "Degree" on 6/8. Additionally, regarding structure metrics comparison, "Local" is comparable to or better than "All" on 5/8 datasets, "Degree" is superior to "All" on 5/8, and "Degree" is comparable to or better than "Local" on 5/8. These results demonstrate that: 1) RE similarity-derived features is indispensable for model performance; 2) degree alone suffices as a structural metric in feature construction, outperforming both global and other local metrics.

\begin{table*}[h]
\caption{Testing of different features.}
\centering
\resizebox{\textwidth}{!}{%
\begin{tabular}{cccccccccc}
\hline
Features &
  Metrics &
  ca-HepTh &
   Facebook 
&Crime &
  Erd\H{o}s&
  Euroroad &
  Hamsterster &
  Sister cities &
  Wikipedia \\ \hline
\multirow{3}{*}{ReEig+Degree}&
  ACC &
  \textbf{0.9704} &
   0.9401 
&0.9488 &
  0.8847&
  0.7828 &
  \textbf{0.9616}&
  \textbf{0.9868} &
  \textbf{0.9694} \\
 &
  NMI &
  \textbf{0.8267} &
   0.7181 
&\textbf{0.7333}&
 0.5135 &
  0.2774&
  \textbf{0.7921}&
  \textbf{0.9058} &
  \textbf{0.8229} \\
 &
  ARI &
  \textbf{0.8840} &
   0.7735 
&0.8032 &
  0.5809 &
  0.3156&
  \textbf{0.8509}&
  \textbf{0.9474} &
  \textbf{0.8803} \\ \hline
\multirow{3}{*}{Degree} &
  ACC &
  0.9656 &
   0.9434
&\textbf{0.9504} &
  0.8774&
  0.7748 &
  0.9521 &
  0.9862 &
  0.9604 \\
 &
  NMI &
  0.8058 &
   0.7281
&0.7176&
  0.4890 &
  0.2666 &
  0.7567 &
  0.9023 &
  0.7872 \\
 &
  ARI &
  0.8663 &
 0.7851
&\textbf{0.8089} &
  0.5546 &
  0.2979 &
  0.8157 &
  0.9449 &
  0.8467 \\ \hline
\multirow{3}{*}{ReEig+All}&
  ACC & 0.9522 & \textbf{0.9621} & 0.9421 & \textbf{0.8917} & 0.7881 & 0.9588 & 0.9092 & 0.9604 \\
& NMI & 0.7300 & \textbf{0.7935 }& 0.7249 & 0.5384 & 0.2554 & 0.7809 & 0.6000 & 0.7872 \\
& ARI & 0.8153 &\textbf{ 0.8528} & 0.7794 & \textbf{0.6036 }& 0.3266 & 0.8404 & 0.6747 & 0.8467 \\ \hline

\multirow{3}{*}{All}&
  ACC & 0.9406 & 0.9541 & 0.9438 & 0.8913 & 0.7616 & 0.9281 & 0.9742 & 0.9604 \\
& NMI & 0.6715 & 0.7640 & 0.7162 & \textbf{0.5507} & 0.1943 & 0.6484 & 0.8329 & 0.7872 \\
& ARI & 0.7718 & 0.8234 & 0.7852 & 0.6008 & 0.2600 & 0.7336 & 0.8984 & 0.8467 \\ \hline

\multirow{3}{*}{ReEig+Local}&
  ACC & 0.8071 & 0.9567 & 0.9421 & 0.8793 & 0.7497 & 0.9613 & 0.9784 & 0.9604 \\
& NMI & 0.2667 & 0.7735 & 0.7249 & 0.5191 & 0.2230 & 0.7903 & 0.8494 & 0.7872 \\
& ARI & 0.3715 & 0.8331 & 0.7794 & 0.5594 & 0.2451 & 0.8496 & 0.9144 & 0.8467 \\ \hline

\multirow{3}{*}{Local}&
  ACC & 0.9313 & 0.9551 & 0.9488 & 0.8807 & \textbf{0.8079} & 0.9491 & 0.9658 & 0.9604 \\
& NMI & 0.6256 & 0.7675 & 0.7191 & 0.5180 & \textbf{0.2860} & 0.7115 & 0.7795 & 0.7872 \\
& ARI & 0.7418 & 0.8270 & 0.8030 & 0.5645 & \textbf{0.3736} & 0.8046 & 0.8663 & 0.8467 \\ \hline
\multirow{3}{*}{ReEig} &
  ACC &
  0.8480 &
   0.9368 
&0.8017 &
  0.8431 &
  0.6781 &
  0.9476 &
  0.7110 &
  0.9604 \\
 &
  NMI &
  0.3521 &
   0.6493 
&0.2739 &
  0.3887 &
  0.0342 &
  0.6875 &
  0.0995 &
  0.7872 \\
 &
  ARI &
  0.4790 &
   0.7583 &0.3724 &
  0.4474 &
  0.0540 &
  0.7986 &
  0.1726 &
  0.8467 \\ \hline
\end{tabular}%
}
\label{tab:test_fea}
\end{table*}

\subsubsection{Testing of $k_p$ and $k_n$} \label{subsubsec:ablation-kpn}
\begin{figure*}
    \noindent
     \begin{tabular}{@{}cccc@{}}
        \includegraphics[width=0.23\textwidth]{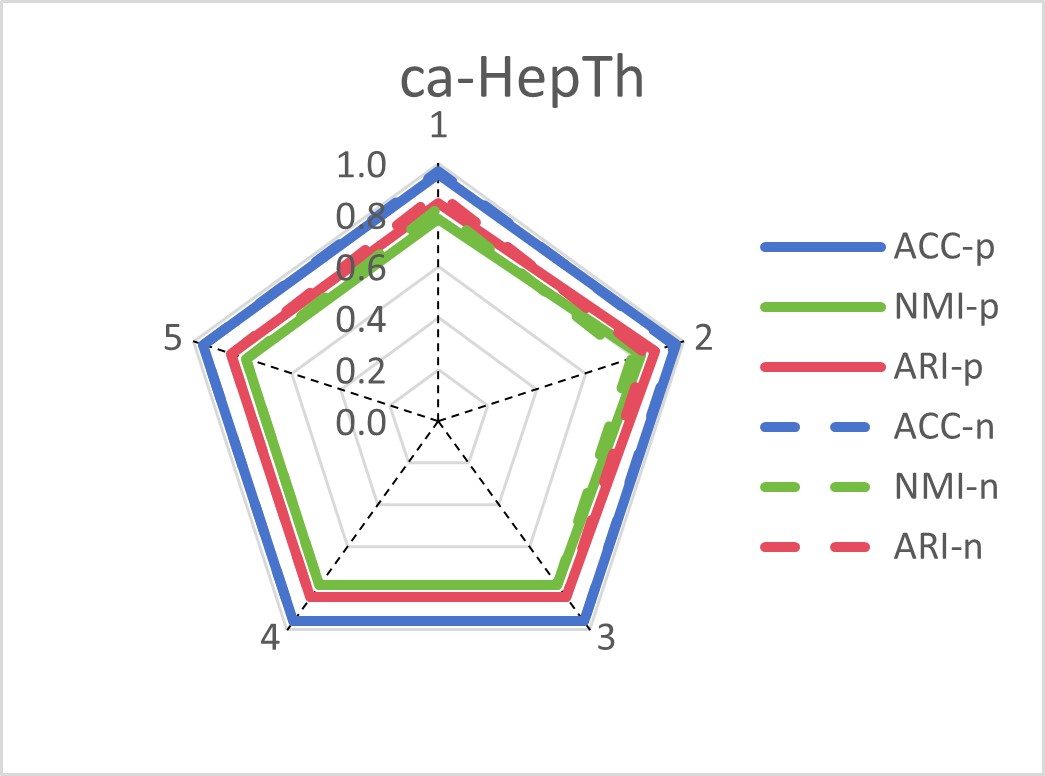}&
        \includegraphics[width=0.23\textwidth]{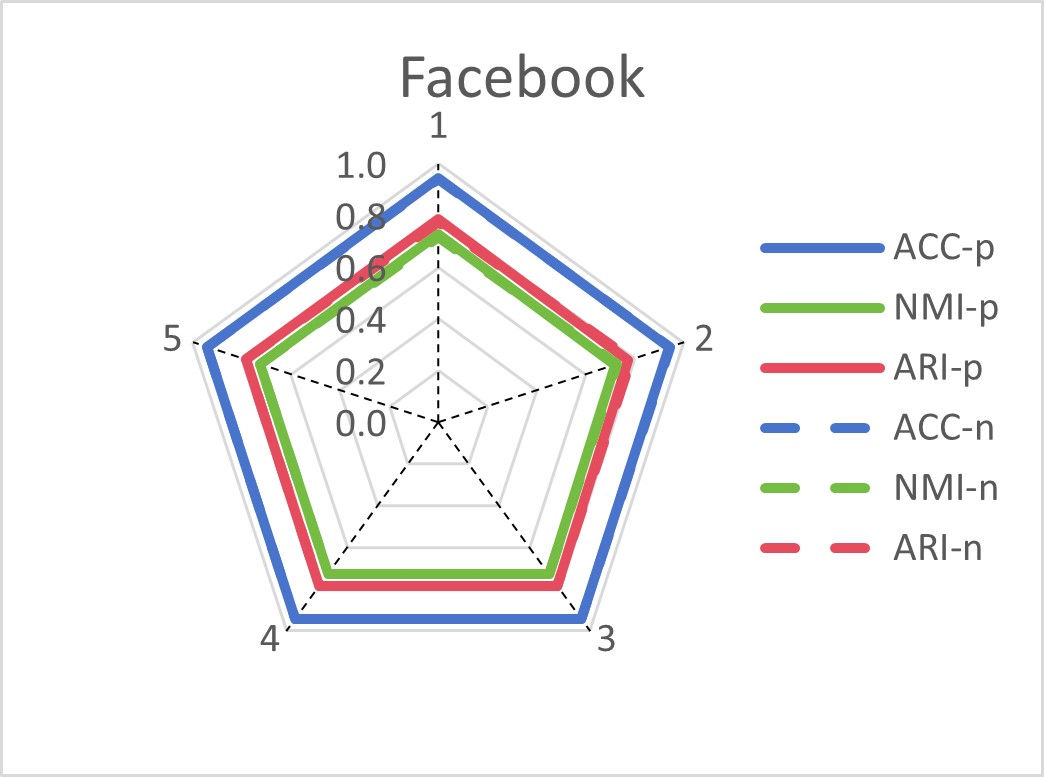}&
        \includegraphics[width=0.23\textwidth]{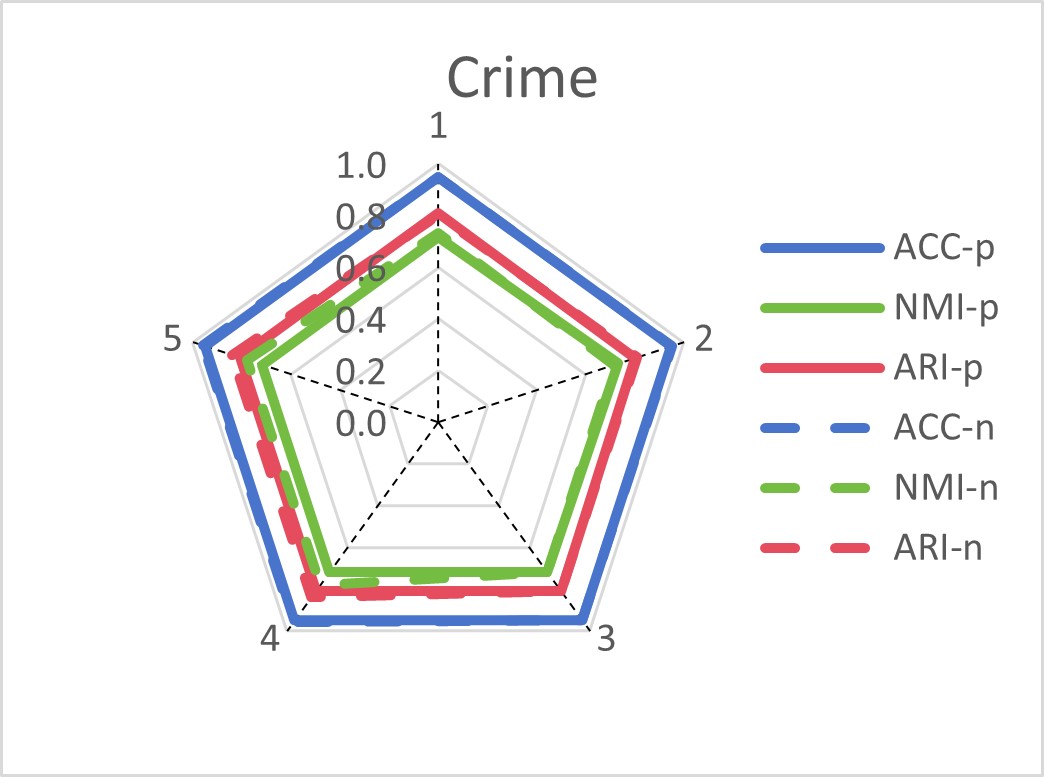}&
        \includegraphics[width=0.23\textwidth]{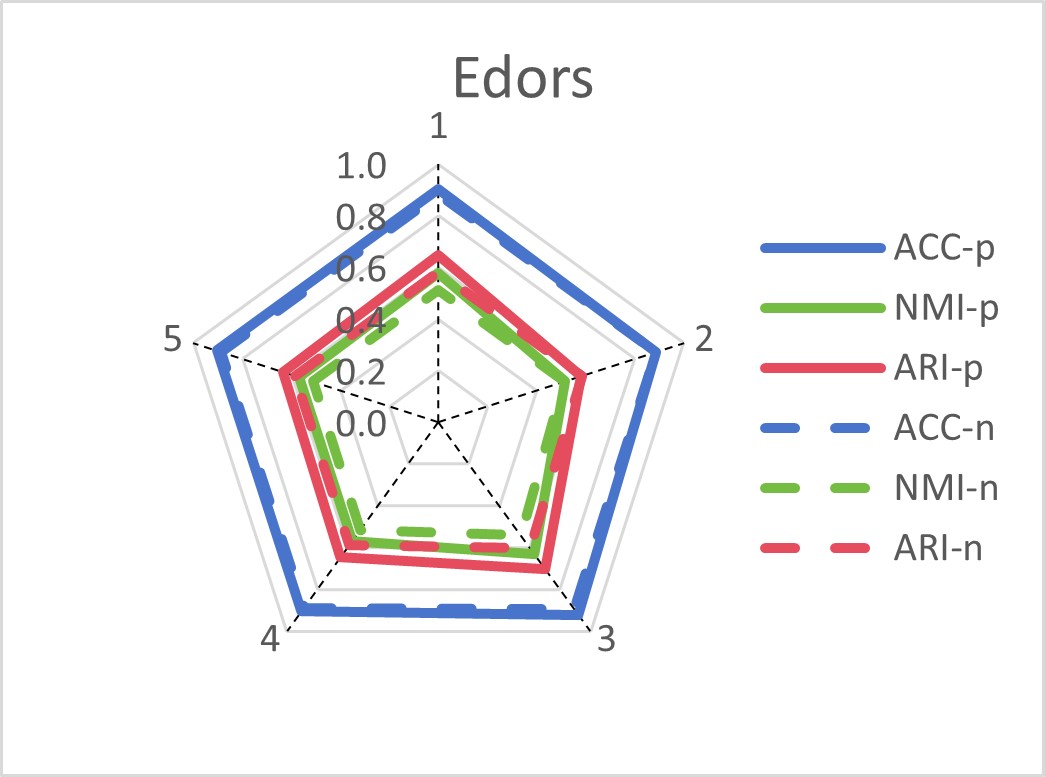}\\
        \includegraphics[width=0.23\textwidth]{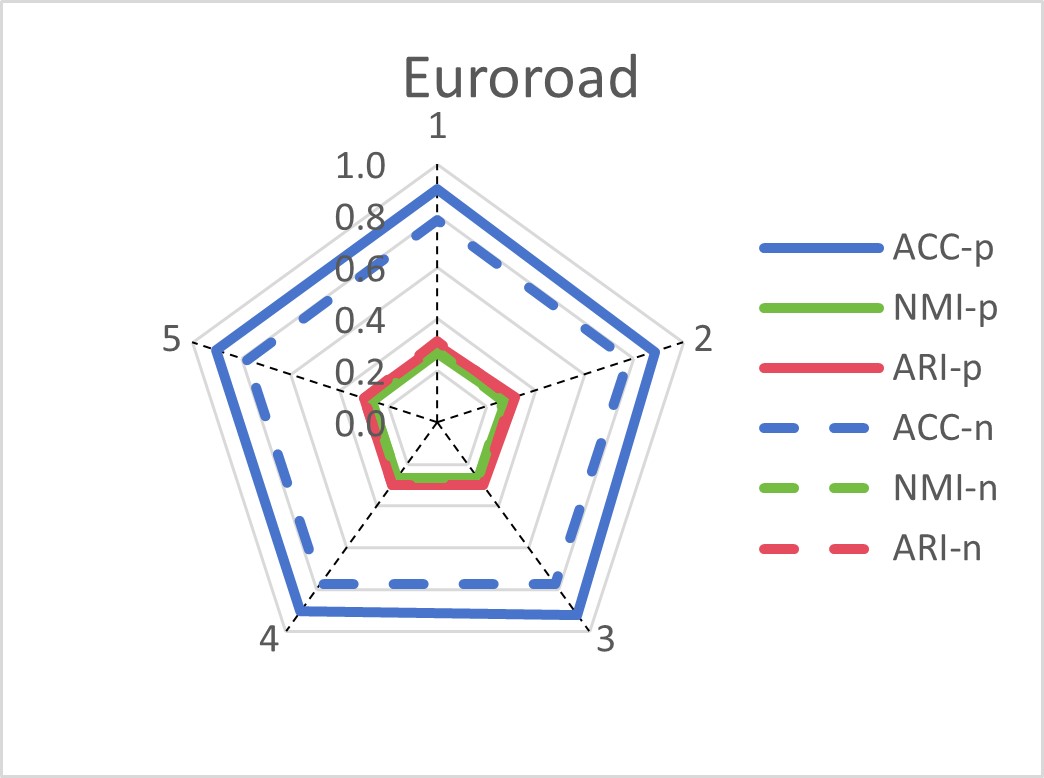}&
        \includegraphics[width=0.23\textwidth]{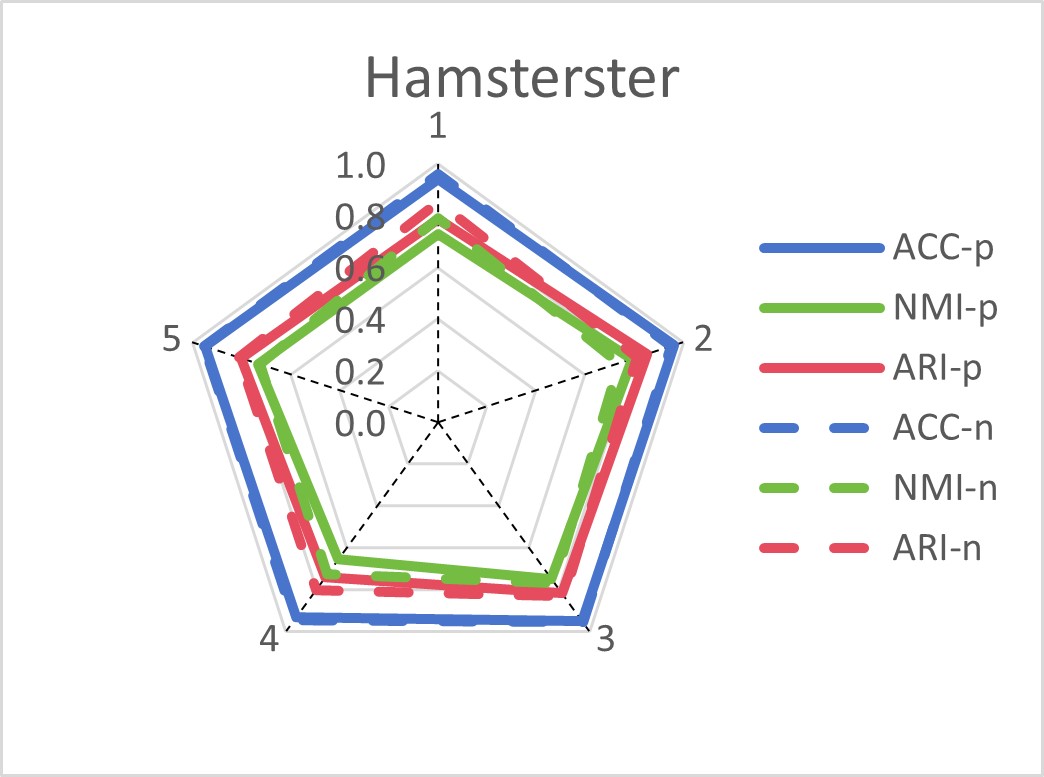}&
        \includegraphics[width=0.23\textwidth]{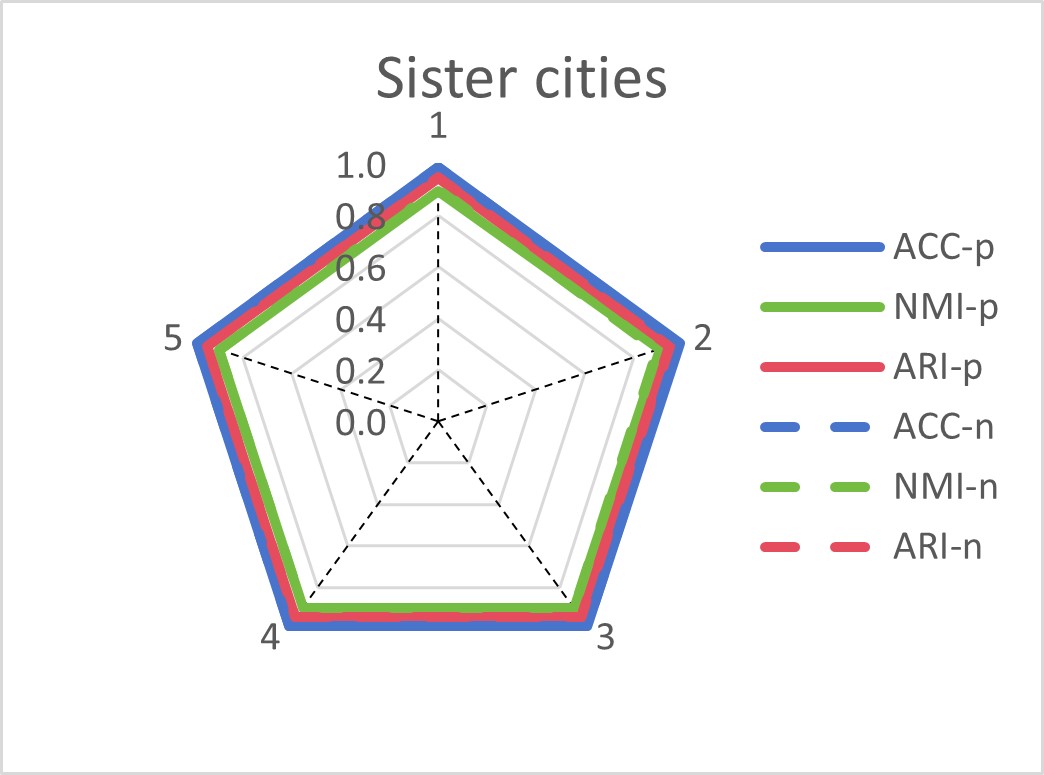}&
        \includegraphics[width=0.23\textwidth]{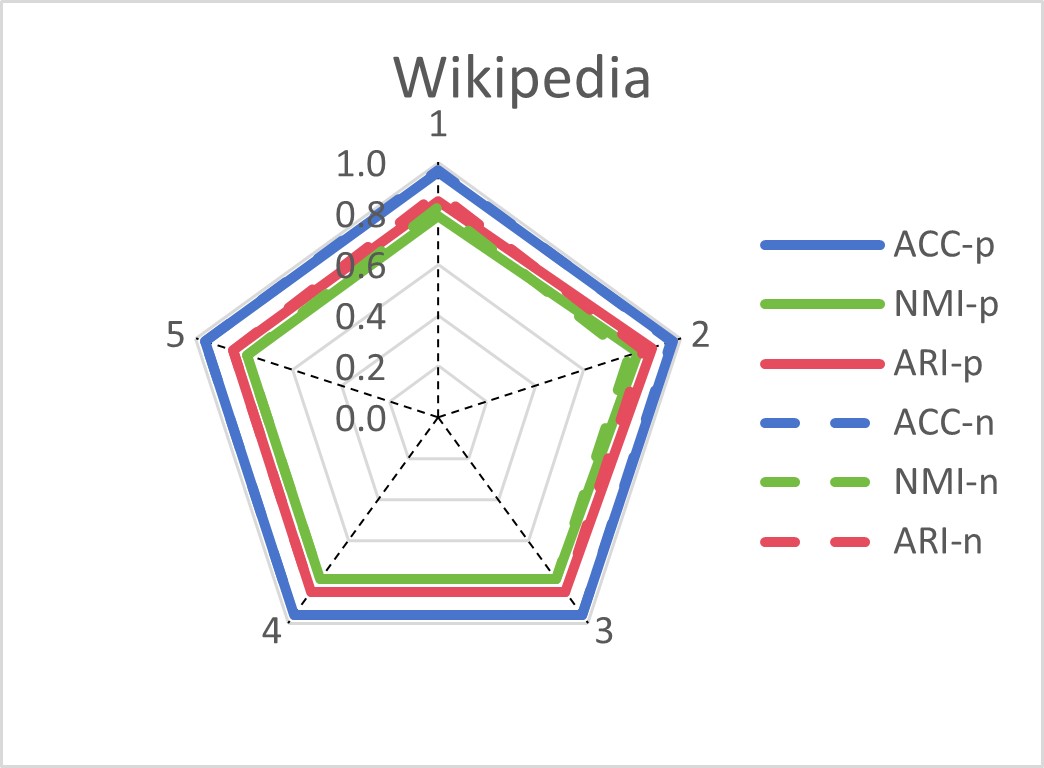}
    \end{tabular}
    \caption{The results of ReCC with different numbers of positive and negative samples.}
    \label{fig:k_pn}
\end{figure*}

ReContrastive in ReCC employs two parameters for contrastive learning: the number of positive samples $k_p$ and the number of negative samples $k_n$. To evaluate their impact, we vary each parameter within $[1, 5]$ (with interval of 1) while fixing the other at its default value. As shown in Fig. \ref{fig:k_pn}, the performance curves for $k_p$ (from 1 to 5) approximate a regular pentagon across all evaluation metrics (ACC, NMI, ARI) on all datasets, except Erd\H{o}s and Hamsterster. On these two datasets, the NMI and ARI curves slightly deviate from this pattern, with optimal performance at $k_p = 3$, $k_n = 3$ on Erd\H{o}s, and $k_p = 2$, $k_n = 1$ on Hamsterster). A similar trend holds for $k_n$. These results indicate ReCC's stability across the tested parameter ranges.

\section{Conclusion}
In this work, we propose ReCC, a RE deep contrastive clustering framework, for unsupervised influential node identification in complex networks. Unlike previous deep learning methods that rely heavily on labeled data, ReCC reformalizes this task as a clustering problem, and provides a novel unsupervised framework to distinguish influential and non-influential nodes without the requirement of labels. ReCC begins by employing GCN to encode nodes into latent embeddings, which serve as the foundation for subsequent clustering. To enhance the discriminative power of these embeddings, ReCC introduces ReContrastive, a contrastive learning mechanism that leverages RE similarity to generate positive and negative samples for contrastive learning. This approach eliminates the need for multi-embedding generation, which is typically required by conventional contrastive learning mechanisms. In addition, ReCC enriches node representations by combining RE similarity-derived features with structural metrics, resulting in more expressive features that better capture node properties. The training process consists of two phases: an initial pre-training using network reconstruction loss, and a fine-tuning where the model is jointly optimized via the ReContrastive loss and KL-divergence loss to refine clustering performance. Extensive experiments demonstrate that ReCC outperforms state-of-the-art clustering models in unsupervised influential node identification across diverse datasets. The success of ReCC can be largely attributed to ReContrastive, which serves as a critical component in enhancing the model’s discriminative ability. Moreover, ReContrastive is significantly more efficient than existing graph contrastive learning mechanisms, owing to its use of fewer positive and especially negative samples per node. Furthermore, ReCC maintains stable performance across a relatively wide range of positive and negative sample sizes, demonstrating notable robustness to these hyperparameters. The results also highlight the effectiveness of combining RE similarity-derived features with structural metrics to enrich node representation.

ReCC offers an effective solution for influential node identification in label-scarce real-world scenarios. However, ReCC dichotomizes nodes as influential or non-influential without accounting for finer-grained influence scores. Further research is needed to evaluate the validity of its multi-clustering extensions for applications requiring influence score stratification.














\appendix

\section{The analysis of structural metrics} \label{sec:appendix}
Numerous structural metrics have been proposed in the literature. Table \ref{tab:global-metrics} lists four widely used global structural metrics that measure a node’s structural position based on the entire network, while Table \ref{tab:node-level} presents eleven local structural metrics that rely solely on a node’s neighborhood. Despite being defined from different perspectives, these metrics may not perform entirely differently in measuring a node’s structural position. For instance, some metrics might capture similar structural positions despite having distinct definitions. To explore potential correlations among these metrics, we analyzed their relationships across dozens of networks from various domains, with node counts ranging from 849 to 1,134,890.

\begin{figure*}
    \centering
    \includegraphics[width=0.5\textwidth]{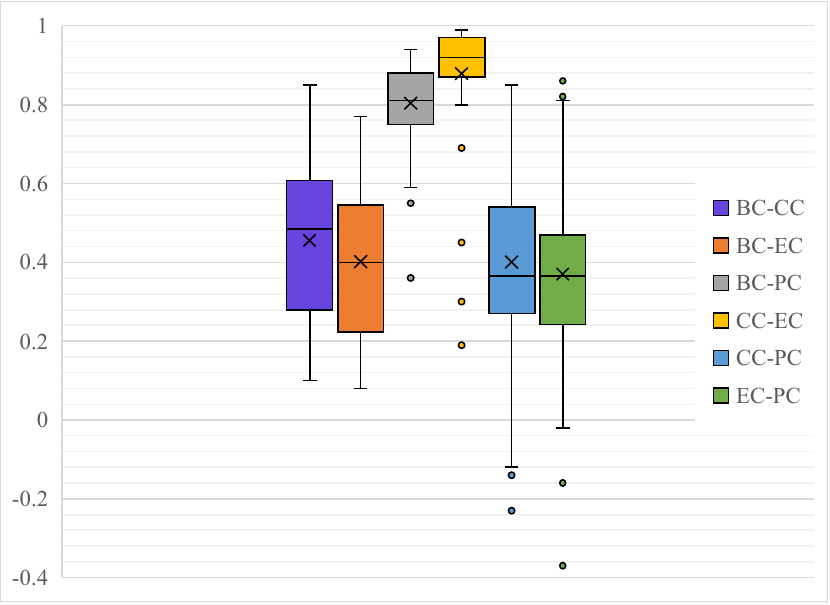}
    \caption{The statistical results on correlation values for each global-metric pair across all networks: the whiskers mark the best and worst cases, $\times$ indicates the mean, the box’s midline the median, and dots the outliers.}
    \label{fig:s1-global-correlation}
\end{figure*}

\begin{figure*}[t]
  \centering
  \begin{minipage}[t]{0.48\linewidth}
     \centering
    \includegraphics[width=\textwidth]{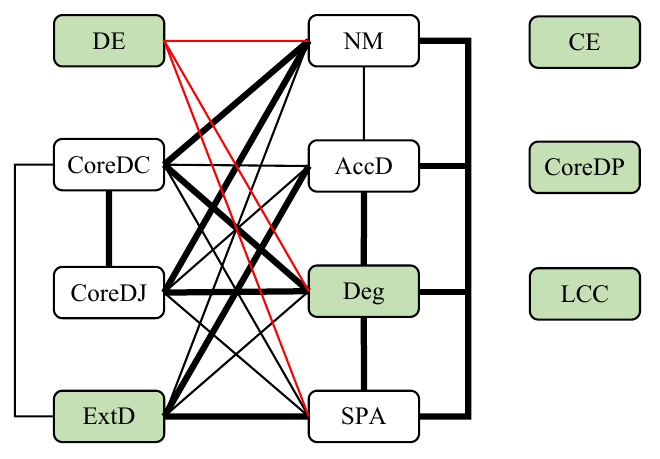}
    \label{fig:local-correlation1}
  \end{minipage}\hfill   
  \begin{minipage}[t]{0.48\linewidth}
    \centering
    \includegraphics[width=\textwidth]{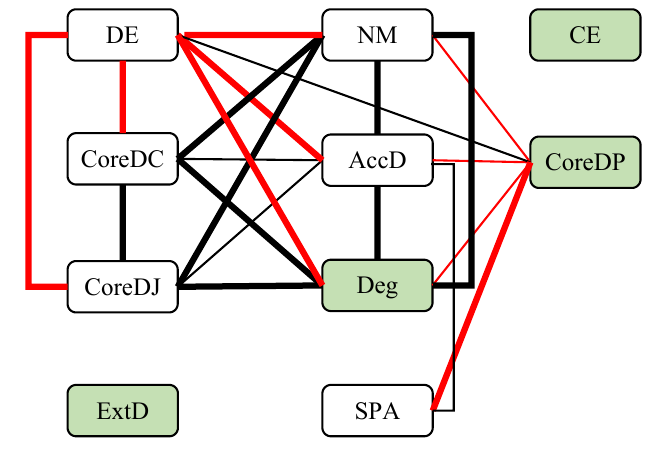}
    \label{fig:local-correlation2}
  \end{minipage}
  \caption{The correlations between local metrics with mean $\geq 0.80$ and standard deviation $\leq 0.10$ (the left subfigure corresponds to the seven star-structured networks, while the right subfigure represents the remaining networks). Red edges denote negative correlations, black edges positive correlations; line width indicates strength ($> 0.90$ thick, $0.80-0.90$ thin). Local metrics in green denote representatives.}
  \label{fig:local-correlation}
\end{figure*}

For global metrics we computed pairwise Spearman correlations between the full-network rankings. For local metrics we first obtained, for each node, its neighborhood rank: the count of neighbors whose metric values exceed its own, normalized by degree. Pearson correlations were then calculated on these normalized neighborhood ranks. This procedure yields two findings:
\begin{itemize}
    \item 1) Among global metrics, Closeness and Eigenvector are strongly correlated, as are Betweenness and PageRank (see Fig. \ref{fig:s1-global-correlation}); thus selecting one from each pair suffices as representative global metrics.
    \item 2) For the local metrics, Figure \ref{fig:local-correlation} displays their correlations. The left subfigure shows the results for seven star-structured networks (all nodes have an LCC value of 0), which are analyzed separately. The right subfigure shows the correlations across all remaining networks. The patterns indicate that Deg, ExtD, DE, CE, LCC, and CoreDP suffice as representative local metrics.
\end{itemize}

\begin{table*}[h]
\centering
\caption{Definitions of global structural metrics.}
\label{tab:global-metrics}
\footnotesize
\begin{tabular}{@{}p{1.7cm}p{7cm}l@{}}
\toprule
\textbf{Metrics} & \textbf{Characteristics of a node} & \textbf{Formulas} \\ \midrule
Eigenvector (EC) & Connected to many other nodes and/or other high-degree nodes. &
$\displaystyle EC_i = \frac{1}{\lambda_1}\sum_{j}A_{ji}v_j$ \\[4mm]

PageRank (PC) & Connected to many other nodes and other high-degree nodes. &
$\displaystyle PR_i = \beta + \alpha\sum_{j}\frac{A_{ji}v_j}{k_j}$ \\[4mm]

Closeness (CC) & Low average shortest path length to other nodes in the network. &
$\displaystyle CC_i = \frac{N-1}{\sum_{j}l_{ij}}$ \\[4mm]

Betweenness (BC) & Lies on many shortest topological paths linking other node pairs. &
$\displaystyle BC_i = \sum_{p\neq i,p\neq q,q\neq i}\frac{g_{pq}(i)}{g_{pq}}$ \\[4mm]
\bottomrule
\end{tabular}

\end{table*}

\begin{table*}[htbp]
\centering
\caption{Definitions of local structural metrics ($N^{+}(u)$ is the combination of $N(u)$ and $u$).}
\label{tab:node-level}
\resizebox{\linewidth}{!}{
\begin{tabular}{@{}p{2cm}p{5cm}l@{}}
\toprule
\textbf{Metrics} & \textbf{Definitions} & \textbf{Formulas} \\ \midrule
Degree (Deg)               & The number of neighbors for the vertex. & $D(u)=|N_{u}|$ \\
Extended degree (EXTD)            & The number of edges attached to the node plus the number of edges attached to each of the node's neighbors. & $EXTD(u)= d_{u}+\sum_{v\in N(u)}d_{v}$ \\
Accumulated degree (ACCD)           & Estimate the impact based on not only the degree of its direct neighbors, but also the degree of the neighbors of these direct neighbors. & $ACCD(u)= d_u+\sum_{v\in N(u)}(d_{v}+\sum_{w\in N(v)}d_{w})$ \\
Node Mass (NM)              & The number of links in the neighborhood community of a node. & $NM(u)=\left|\left\{(v,w)\in\mathcal{E}\mid v,w\in N^{+}(u)\right\}\right|$ \\
Conductance of egonet (CE)              & Evaluates the density level of egonet of node $i$ by counting the edges in ego($i$) and the ones between ego($i$) and the remained network. & $CE(u)=\dfrac{\left|\left\{(v,w)\in\mathcal{E}\mid v\in N^{+}(u),w\in\mathcal{V}-N^{+}(u)\right\}\right|}{\min\!\bigl(\text{vol}(N^{+}(u)),\text{vol}(\mathcal{V}-N^{+}(u))\bigr)}$ \\
Density of egonet (DE)              & Evaluates the density of ego($i$) by comparing with a complete graph induced by the same nodes. & $DE(u)=\dfrac{\left|\left\{(v,w)\in\mathcal{E}\mid v,w\in N^{+}(u)\right\}\right|}{\left|\left\{(v,w)\mid v,w\in N^{+}(u)\right\}\right|}$ \\
Local clustering coefficient (LCC)             & Evaluates the density of ego($i$) by counting the number of connected neighbor nodes. & $LCC(u)=\dfrac{\left|\left\{(v,w)\in\mathcal{E}\mid v,w\in N(u)\right\}\right|}{\left|\left\{(v,w)\mid v,w\in N(u)\right\}\right|}$ \\
Core dominance Cosine (CoreDC)          & Core dominance measure calculated by cosine index. & $COREDC(u)=\sum_{v\in N(u)}\dfrac{|N(u)\cap N(v)|}{\sqrt{|N(u)|\,|N(v)|}}$ \\
Core dominance Jaccard (CoreDJ)          & Core dominance measure calculated by Jaccard index. & $COREDJ(u)=\sum_{v\in N(u)}\dfrac{|N(u)\cap N(v)|}{|N(u)\cup N(v)|}$ \\
Core dominance Pearson (CoreDP)          & Core dominance measure calculated by Pearson correlation coefficient. & $COREDP(u)=\dfrac{\sum_{w}(A_{u,w}-\bar{A}_{u})(A_{v,w}-\bar{A}_{v})}{\sqrt{\sum_{w}(A_{u,w}-\bar{A}_{u})^{2}}\sqrt{\sum_{w}(A_{v,w}-\bar{A}_{v})^{2}}}$ \\
Core dominance PA(SPA)         & Core dominance measure calculated by Preference Attachment. & $SPA(u)=\sum_{v\in N(u)}d_{u}\,d_{v}$ \\
\bottomrule
\end{tabular}
}
\end{table*}



\section*{Acknowledgments}
The work was substantially supported by the National Natural Science Foundation of China under Grant Nos. 61802034, 62472195, 62076109, 623B2041, and 62072115. This work was also supported by Shanghai Science and Technology Innovation Action Plan Project under Grant No. 22510713600. We are grateful for their financial support, which made this study possible.

\bibliographystyle{unsrt}
\bibliography{sample}

\end{document}